\documentclass[a4paper,fleqn,usenatbib]{mnras}

\usepackage{newtxtext,newtxmath}

\usepackage[T1]{fontenc}
\usepackage{ae,aecompl}

\usepackage{graphicx}	
\usepackage{amsmath}	
\usepackage{multirow}

\usepackage{bm}
\usepackage{enumerate}
\usepackage{xcolor}

\title[Magnification of GWs from stellar-mass BBHs]{Lensing magnification: gravitational waves from coalescing stellar-mass binary black holes}

\author[X. Shan, C. Wei and B. Hu]{
Xikai Shan$^{1}$,
Chengliang Wei$^{2}$,
Bin Hu$^{1}$\thanks{E-mail: bhu@bnu.edu.cn}
\\
$^{1}$Department of Astronomy, Beijing Normal University, Beijing 100875, China\\
$^{2}$Purple Mountain Observatory, Chinese Academy of Sciences, Nanjing, 210023, China
}

\date{Accepted XXX. Received YYY; in original form \today}

\pubyear{2021}

\begin{document}
\label{firstpage}
\pagerange{\pageref{firstpage}--\pageref{lastpage}}
\maketitle

\begin{abstract}
Gravitational waves (GWs) may be magnified or de-magnified due to lensing. 
This phenomenon will bias the distance estimation based on the matched filtering technique. 
Via the multi-sphere ray-tracing technique, we study the GW magnification effect and selection effect with particular attention to the stellar-mass binary black holes (BBHs).   
We find that, for the observed luminosity distance $\lesssim 3~\mathrm{Gpc}$, which is the aLIGO/Virgo observational horizon limit, the average magnification keeps as unity, namely unbiased estimation, with the relative distance uncertainty $\sigma(\hat{d})/\hat{d}\simeq0.5\%\sim1\%$. 
Beyond this observational horizon, the estimation bias can not be ignored, and with the scatters $\sigma(\hat{d})/\hat{d} = 1\%\sim 15\%$. 
Furthermore, we forecast these numbers for Einstein Telescope.
We find that the average magnification keeps closely as unity for the observed luminosity distance $\lesssim 90~\mathrm{Gpc}$. 
The luminosity distance estimation error due to lensing for Einstein Telescope is about $\sigma(\hat{d})/\hat{d} \simeq 10\%$ for the luminosity distance $\gtrsim 25~\mathrm{Gpc}$.
Unlike the aLIGO/Virgo case, this sizable error is not due to the selection effect. It purely comes from the unavoidably accumulated lensing magnification.
Moreover, we investigated the effects of the orientation angle and the BH mass distribution models.
We found that the results are strongly dependent on these two components.
\end{abstract}




\section{Introduction}
Since the discovery of GW events in 2015, the advanced LIGO/Virgo consortium \citealt{Acernese_2014,2015} has identified $44$ GWs generated by the merging of (BBHs) during the observation O1, O2 and O3a run \citealt{TheLIGOScientific:2016pea, Abbott:2016blz, Abbott:2016nmj,Abbott:2017oio, TheLIGOScientific:2017qsa, Abbott:2017gyy, Abbott:2017vtc, LIGOScientific:2018mvr, abbott2021gwtc2}. 
The third-generation GW detectors, such as Einstein Telescope and Cosmic Explorer \citealt{2014ASSL..404..333P, Evans:2016mbw}, will detect about $10^5$ GWs each year. 
GWs will become a powerful tool for studying fundamental physics, astrophysics, and cosmology. 
However, the GWs may be magnified (de-magnified) due to the intervening over-dense (under-dense) structures along the line-of-sight. 
Hence, it will lead to an under-estimation (over-estimation) of the luminosity distance or an over-estimation (under-estimation) of the chirp mass \citealt{dai2017waveforms, oguri2018effect, cusin2019strong}. 
These are the intrinsic scatters in the GW luminosity distance estimation. First, the current GW cosmology studies are mainly based on the bright/dark siren approach \citealt{Schutz:1986gp, Guidorzi_2017, PhysRevLett.119.161101, Chen:2017rfc, Wang:2020dkc, Abbott:2019yzh}, in which the distance is one of the essential parameters. 
Second, the lensing error will  be one of the dominant errors for the future GW observations.
Third, via the parameter degeneracies, the lensing effect could also affect the reconstruction of the astrophysical properties of BBHs, such as the masses.
Therefore, the gravitational lensing effect deserves more careful studies.
Actually, there already existed extensive studies on the GW lensing phenomena during the past decades, especially after the GW discoveries in 2015.  
\citealt{wang1996gravitational} firstly predicted the strongly lensed events rate of GWs from the mergers of binary neutron stars under the advanced LIGO (aLIGO) configuration. 
\citealt{Holz:2005df,Hirata:2010ba} investigated the Gaussian and non-Gaussian statistics of GW lensing magnification.  
\citealt{PhysRevD.90.062003, dai2017waveforms} discussed the GW waveform distortion caused by lensing. 
\citealt{Smith:2018gle, Ng:2017yiu, li2018gravitational, oguri2018effect, Smith_2018, Robertson:2020mfh} studied the lensing rate for aLIGO and Einstein Telescope (ET). 
\citealt{Smith:2018kbc, yu2020strong, Hannuksela:2020xor, Ryczanowski:2020mlt, Sereno:2011ty} proposed to use the strong lensing of GWs to improve the localization of GW events. 
\citealt{hannuksela2019search, Li:2019osa, McIsaac:2019use, Pang:2020qow, dai2020search, Abbott:2020mjq, Liu:2020par}  searched for the lensed GWs in aLIGO/Virgo data.
There are many more relevant works, such as \citealt{PhysRevLett.118.091101,PhysRevLett.118.091102,Smith_2018,haris2018identifying,Zhao:2019gyk,Contigiani_2020,Mukherjee:2020tvr}, which we will not introduce in detail here. 

Because of the lensing magnification and related selection effects, the GW luminosity distance estimation can be biased and scattered. 
There are pioneer works focusing on this topic \citealt{cusin2019strong,hannuksela2019search,Cusin:2020ezb}. 
Especially, \citealt{cusin2019strong} recently discussed the average GW magnification from the stellar-mass binary black holes by assuming a single lens plane.  
In this paper, we extend this calculation by including the multiple lenses effects via the ray-tracing technique developed by \citealt{Wei:2018uwb}. 

The rest of this paper is structured as follows. 
In Section~\ref{sec:model}, we introduce the theoretical framework, including the GW distribution model as well as the luminosity distance estimator. 
In Section~\ref{num_results}, we present the average magnification and the distance estimation uncertainties for aLIGO/Virgo O1 + O2 + O3a run. 
In Section~\ref{for_et}, we give the forecast for ET.  
Finally, we summarize and discuss our result in Section~\ref{sec:summary}. 
In this work, we assume the Planck $\Lambda$CDM cosmology~\citealt{collaboration2018planck}. 

\section{Methodology}
\label{sec:model}

In order to take the multiple lenses effect into account, we calculate the GW lensing magnification by using the ray-tracing method~\citealt{Wei:2018uwb} based on N-body simulation which is part of the ELUCID project~\citealt{Wang:2014hia,Li:2016nbw,Wang:2016qbz,Tweed:2017wsu}.
In detail, the particle masses in the adopted N-body simulation are $3.4\times10^8 M_{\odot}$.
Furthermore, our targeted GW frequency ranges are in $10\sim1000~{\rm Hz}$. 
As demonstrated in~\citealt{takahashi2003wave,li2018gravitational}, when the lens masses are greater than $10^5~M_{\odot}(f/Hz)^{-1}$, where $f$ is the GW frequency, one can apply the geometric optics approximation.  
Hence, it is safe to neglect the wave optics in the following calculation. 
The rest of this section is scheduled in the following manner. 
In Section~\ref{avemag}, we list the formulas to calculate the average magnification. 
In Section~\ref{GWM}, we present the compact binary population model. 
Finally, we introduce the estimator with the selection effect and the distance estimation uncertainty in Section~\ref{Est}.

\subsection{Average magnification}
\label{avemag}
In this work, we are interested in the average magnification as well as the scatters of the estimated GW luminosity distance. 
This is complementary to the galaxy surveys because GW observation directly measures the luminosity distance instead of redshift.  

For the stationary noise case, the signal-to-noise ratio (SNR) $\rho$ is defined as~\citealt{flanagan1997measuring}
\begin{align}
\label{snr}
		\rho^{2}=4 \int_{0}^{\infty} \mathrm{d} f \frac{|h(f)|^{2}}{S_{n}(f)}\;,
\end{align}
where $S_{n}(f)$ is the sensitivity curve of detectors and $h(f)$ is the GW strain in the frequency domain. 
To obtain the SNR for a specific interferometer, in theory, it is necessary to know the orientation of the detector with respect to the BBH and inclination. 
This angle information can be encapsulated into the orientation function $\Theta^2$ (see~\citealt{finn1993observing} for more details).
Here, we consider a general configuration ($\Theta^2$ = 64/25) and an optimal configuration ($\Theta^2 = 16$), respectively.

Given a GW detection threshold ($\rho_\mathrm{lim}$), the number of GW events located in the redshift bin $(z_s\pm\mathrm{d}z_s/2)$ is 
\begin{align}
	\mathrm{d} \mathcal{N}\left(\rho_{\mathrm{lim}}, z_{s}\right)=\mathrm{d} z_{s} \int_{\rho_\mathrm{lim}}^{\infty} n\left(\rho, z_{s}\right) \mathrm{d} \rho \;,
	\label{eq:dN}
\end{align}
where $n\left(\rho, z_{s}\right)$ is the GW events number density in the SNR and redshift space. 
We emphasize that, here, we have not yet considered the lensing magnification.  
After passing through the lenses, the observed strain $h(f)$ is magnified (or de-magnified) by a factor $\sqrt{\mu}$ 
\begin{align}
	h_\mathrm{lens}(f) = \sqrt{\mu} h_\mathrm{unlens}(f) \ .
\end{align}
After including magnification, Eq. (\ref{eq:dN}) shall be written as
\begin{align}
\label{source_number}
	\mathrm{d} \mathcal{N}\left(\rho_{\mathrm{lim}}, z_{s}\right)=\mathrm{d} z_{s} \int_{\mu_\mathrm{min}}^{\mu_\mathrm{max}} \mathrm{d} \mu p\left(\mu, z_{s}\right) \int_{\rho_{\mathrm{lim} / \sqrt{\mu}}}^{\infty} n\left(\rho, z_{s}\right) \mathrm{d} \rho \;,
\end{align}
where $p\left(\mu, z_{s}\right)$ is the magnification probability distribution function (PDF).
For redshifts between $1$ and $20$, we directly read the public data released by \citealt{Takahashi:2011qd}. 
However, most of the current aLIGO/Virgo BBH events are localized within the luminosity distance radius $3000$ Mpc (or $z\lesssim 0.5$).  
For this reason, we make use of the full-sky multi-sphere ray-tracing code developed by \citealt{Wei:2018uwb} to get the PDF for $z\lesssim 1$. 
In detail, we sample the lens sphere uniformly in the comoving distance with the separation of $100h^{-1}$ Mpc.  

Notice that, in Eq.~(\ref{source_number}), magnification will lower the detection SNR threshold by a factor $1/\sqrt{\mu}$.
Moreover, the magnification upper (lower) limit $\mu_\mathrm{max} (\mu_\mathrm{min})$ refers to the maximum (minimum) limit of $\mu$ in the numerical ray-tracing result. 

\begin{figure}
	\centering 
	\includegraphics[width=\columnwidth]{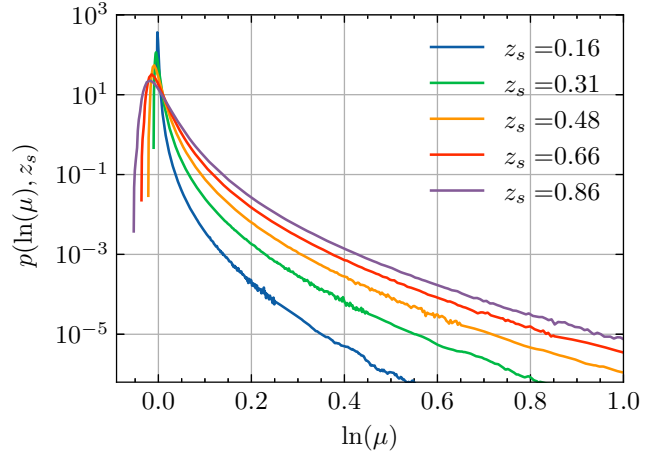}
	\caption{Lensing magnification PDF as a function of $\ln\mu$. 
	Different colors represent different source redshifts $z_s$.}
	\label{fig:pdf}
\end{figure}

Figure~\ref{fig:pdf} shows the magnification PDF as a function of $\ln\mu$ at different source redshifts $z_s$.
One can see that the maximum value of PDF will decrease and shift left with the increase of $z_s$.
Additionally, the de-magnification effect (corresponding to $\ln\mu<0$) becomes essential at higher redshift.
Now, we can calculate the mean magnification of a GW event located at redshift $z_s$
\begin{align}
\label{ave_mag_z}
	\langle\mu\rangle\left(z_{s}\right)=\frac{\int_{\mu_\mathrm{min}}^{\mu_\mathrm{max}} \mathrm{d} \mu \mu p\left(\mu, z_{s}\right) \int_{\rho_{\mathrm{lim} / \sqrt{\mu}}}^{\infty} n\left(\rho, z_{s}\right) \mathrm{d} \rho}  {\int_{\mu_\mathrm{min}}^{\mu_\mathrm{max}} \mathrm{d} \mu p\left(\mu, z_{s}\right) \int_{\rho_{\mathrm{lim} / \sqrt{\mu}}}^{\infty} n\left(\rho, z_{s}\right) \mathrm{d} \rho} \ .
\end{align}
In this formula, we take the observational sensitivity aspects into account. The resulted average magnification may deviate from unity. 
This is the selection effect.  
It shows in our calculations via the SNR which appears at the lower limit of integral Eq.~(\ref{source_number}).
Since the redshift of GW event can not be measured directly, Eq.~(\ref{ave_mag_z}) is only applicable to the case where the redshift can be obtained by other means, e.g. through the associated electromagnetic counterparts. 
 
For GW observations only, a more naturalistic representation is to show the average magnification in the luminosity distance space~\citealt{cusin2019strong}. 
Hence, we rewrite Eq.~(\ref{source_number}) as
\begin{align}
\begin{split}
	\mathrm{d} \mathcal{N}\left(\rho_{\mathrm{lim}}, d_{\mathrm{obs}}\right)=\mathrm{d} d_{\mathrm{obs}} \int_{\mu_\mathrm{min}}^{\mu_\mathrm{max}} \mathrm{d} \mu p\left(\mu, d_{\mathrm{obs}}\right) \frac{\mathrm{d} z_s}{\mathrm{d} d_{\mathrm{obs}}} \times \\ \int_{\rho_{\mathrm{lim} / \sqrt{\mu}}}^{\infty} n\left(\rho, d_{\mathrm{obs}}\right) \mathrm{d} \rho \ ,
\end{split}
\end{align}
where $d_{\mathrm{obs}}$ is the observed luminosity distance of a GW source. 
The relationship of the ``observed'' and ``true'' luminosity distance is
\begin{align}
	d_{\mathrm{obs}}\left(z_{s}, \mu\right) \equiv \frac{d\left(z_{s}\right)}{\sqrt{\mu}} \ ,
\end{align}
where $d\left(z_{s}\right)$ is the ``true'' luminosity distance without the lensing effect. 
Therefore, the average magnification as a function of $d_\mathrm{obs}$ can be written as
\begin{align}
\label{ave_obs}
	\langle\mu\rangle\left(d_\mathrm{obs}\right) = \frac{ \int_{\mu_\mathrm{min}}^{\mu_\mathrm{max}} \mathrm{d} \mu \mu p\left(\mu, d_{\mathrm{obs}}\right)  \times \mathrm{d} \mathcal{N^\prime}\left(\mu, d_{\mathrm{obs}}\right)} { \int_{\mu_\mathrm{min}}^{\mu_\mathrm{max}} \mathrm{d} \mu p\left(\mu, d_{\mathrm{obs}}\right) \times \mathrm{d} \mathcal{N^\prime}\left(\mu, d_{\mathrm{obs}}\right)} \;,
\end{align}
where 
\begin{align}
\label{mathcal_N}
	\mathrm{d} \mathcal{N^\prime}\left(\mu, d_{\mathrm{obs}}\right) \equiv \mathrm{d} d_{\mathrm{obs}} \frac{\sqrt{\mu}}{d_{s}^{\prime}(z_s)} \int_{\rho_{\mathrm{lim} / \sqrt{\mu}}}^{+\infty} n\left(\rho, d_{\mathrm{obs}}\right) \mathrm{d} \rho \;,
\end{align}
and $d_{s}^{\prime}\left(z_{s}\right) \equiv \mathrm{d} d\left(z_{s}\right) / \mathrm{d} z_s$. 

For convenience, we define the magnification PDF with selection effect as
\begin{align}
\label{cal_p}
	\mathcal{P}_{d_{\mathrm{obs}}}\left(\mu, d_{\mathrm{obs}}\right) \equiv \mathcal{C} p\left(\mu, d_{\mathrm{obs}}\right) \frac{\mathrm{d} \mathcal{N^\prime}\left(\mu, d_{\mathrm{obs}}\right)}{\mathrm{d} d_{\mathrm{obs}}} \;,
\end{align}
which is first introduced into the GW community by \citealt{Cusin:2020ezb},
where $\mathcal{C}$ is the normalization coefficient. 
For the case in which the redshifts can be obtained by other means, 
one needs to replace $d_{\mathrm{obs}}$ in Eq.~(\ref{cal_p}) with $z_s$ to get $\mathcal{P}_{z_s}(\mu, z_s)$.

\subsection{GW distribution modeling}
\label{GWM}
In order to get the average magnification as a function of redshift or observed luminosity distance, it is necessary to know the GW source distribution. 
In this section, we introduce a phenomenological model of the BBH population.
The recent study \citealt{Abbott:2020gyp} shows that the merger rate of stellar mass BBHs is likely to increase with redshift but no faster than the star-formation rate (SFR) for redshift $z_s\lesssim1$. 
In this paper, in order to simplify and calculate the BBH population at redshift $z_s>1$, we assume the merger rate of BBHs is proportional to SFR~\citealt{cusin2019strong} at any redshift.

First, we use the hybrid SFR described in~\citealt{2003MNRAS.339..312S, vangioni2014impact}
\begin{align}
	\Psi(t(z))=A \frac{e^{b\left(z-z_{m}\right)}}{a-b+b \cdot e^{a\left(z-z_{m}\right)}} \;,
\end{align}
where $A=0.24~M_{\odot} /yr/ \mathrm{Mpc}^{3}$, $z_m = 2.3$, $a = 2.2$ and $b = 1.4$. 
Second, we assume the probability density of time-delay between the star formation and coalescence of the BBHs is~\citealt{Vitale_2019}
\begin{align}
	P_{d}\left(t_{\text {delay }}\right) \propto t_{\text {delay }}^{-1} \ ,
\end{align}
where $t_{\text {delay }}$ ranges from $10~{\rm M}yr$ to $10~{\rm G}yr$~\citealt{Vitale_2019}. 
In this paper, we do not attempt to investigate the influence of different time delay models on the final merger rate. 
This is because, as shown in~\citealt{Vitale_2019}, this effect shall not be the dominant theoretical uncertainty compared with the BH mass model, which we will demonstrate lately.   
For simplicity, we take the ``flat in log'' model~\citealt{Vitale_2019}.

Third, we use two different black hole mass distribution models.
One is the \textbf{POWER LAW + PEAK} model (referred to as ``Model C'' in~\citealt{collaboration2018binary}), which is the most accurate model fitting to GWTC-$2$ data~\citealt{Abbott:2020gyp}.
The complete form of this probability distribution is
\begin{align} 
\label{model_c}
\begin{split}
p\left(m_{1} \right)=&\left[\left(1-\lambda_\text{peak}\right) A m_{1}^{-\alpha} \Theta\left(m_{\max }-m_{1}\right)\right.\\ &\left.+\lambda_\text{peak} B \exp \left(-\frac{\left(m_{1}-\mu_{m}\right)^{2}}{2 \sigma_{m}^{2}}\right)\right] S\left(m_{1}, m_{\min }, \delta m\right) \;,\\
p\left(q \mid m_{1}\right)=& C\left(m_{1}\right) q^{\beta_{q}} S\left(m_{2}, m_{\min }, \delta m\right) \;,
\end{split}
\end{align}
where $m_1$ and $q$ stand for the primary BH mass and the BBH mass ratio $m_2/m_1$, respectively.
$\Theta\left(m_{\max }-m_{1}\right)$ is the Heaviside step function.
$S$ refers to a smoothing function, which is
\begin{align}
\begin{split}
&S\left(m \mid m_{\min }, \delta_{m}\right)=
\\ &\left\{\begin{array}{ll}0 & \left(m<m_{\min }\right) \\ {\left[f\left(m-m_{\min }, \delta_{m}\right)+1\right]^{-1}} & \left(m_{\min } \leq m<m_{\min }+\delta_{m}\right) \\ 1 & \left(m \geq m_{\min }+\delta_{m}\right)\end{array}\right. \;,
\end{split}
\end{align}
where 
\begin{align}
f\left(m^{\prime}, \delta_{m}\right)=\exp \left(\frac{\delta_{m}}{m^{\prime}}+\frac{\delta_{m}}{m^{\prime}-\delta_{m}}\right) \;.
\end{align}
The parameters in Eq.~\ref{model_c} we used are $\lambda_\text{peak} = 0.1$, $\alpha = 2.63$, $m_\text{max} = 86.22$, $m_\text{min} = 4.59$, $\mu_m = 33.07$, $\sigma_m = 5.69$, $\delta_m = 4.82$ and $\beta_q = 1.26$, which are the mean value of the posterior distribution of GWTC-2~\citealt{Abbott:2020gyp}. 
This fit is performed excluding GW190814, which is the BHNS merger candidate. 
The coefficients $A$, $B$ and $C$ ensure that the power-law distribution $m_1^{-\alpha}$, Gaussian distribution and the mass ratio distribution are normalized, respectively.

Another mass distribution model is the \textbf{one-parameter POWER LAW} model~\citealt{Fishbach_2017,Wysocki_2019} (referred to as ``Model A'' in~\citealt{collaboration2018binary}). 
This simplified model had difficulties to explain the high-mass BBH events like GW$190521$~\citealt{Abbott:2020gyp,abbott2021gwtc2}.
We apply it here to test the model dependence of our method.
It assumes that the probability density of the primary BH mass is
\begin{align}
	P\left(m_{1}\right) \propto m_{1}^{-\alpha} \ ,
\end{align}
where $\alpha = 2.35$ and $m_1$ is in $[5~M_{\odot} , 50~M_{\odot}]$~\citealt{collaboration2018binary}. 
The probability density of the secondary BH mass is constant, which ranges from $5~M_{\odot}$ to $m_1$. 

Based on these assumptions, the merger rate of BBHs can be expressed as 
\begin{align}
\begin{split}
\label{sfr_bbh}
	&\frac{\mathrm{d} R}{\mathrm{d} m_{1} \mathrm{d} m_{2} \mathrm{d} z_s}= \epsilon \int \frac {\Psi\left(t(z_\mathrm{delay})\right)}{1+z_\mathrm{delay}} \times \\ &p\left(m_{1}\right) p\left(m_{2} \mid m_1\right) P_{d}\left(t_{\mathrm{delay}}\right) \frac{\mathrm{d} V}{\mathrm{d} z_s }\mathrm{d} t_{\mathrm{delay}} \;,
\end{split}
\end{align}
where $z_\mathrm{delay}$ is the redshift at the star-forming and $\epsilon$ is an unfixed ratio between the SFR and the merging rate of BBH.
In the above formula, the factor $1/(1+z_{\rm delay})$ represents the redshift between the star-forming epoch and the present.

Next, we use the O1 + O2 + O3a run data of aLIGO/Virgo collabortion to calibrate the parameter $\epsilon$. 
The observed GW events from BBH can be written as
\begin{align}
\label{obser_num}
	N_{\mathrm{O1 + O2 + O3a}}=  T_{\mathrm{obs}} \int_{\rho\left(m_{1}, m_{2}, z\right) \geq 8} \frac{\mathrm{d} R}{\mathrm{d} m_{1} \mathrm{d} m_{2} \mathrm{d} z} \mathrm{d} m_{1} \mathrm{d} m_{2} \mathrm{d} z \ ,
\end{align}
where $N_{\mathrm{O1 + O2 + O3a}} = 44$. 
We use the open software package \texttt{PyCBC}\footnote{\url{https://pycbc.org/}}~\citealt{2005PhRvD..71f2001A, 2012PhRvD..85l2006A, Canton:2014ena, Nitz:2017svb} to calculate the SNR. 
Furthermore, plug it into Eq.~(\ref{obser_num}) to get $\epsilon$. 
Finally, we can get the number of GW sources $n(\rho,z_s)$ per unit redshift and per unit SNR defined in Eq.~(\ref{eq:dN}) 
\begin{align}
	n(\rho,z_s)=  T_{\mathrm{obs}} \frac{1}{\mathrm{d}\rho}\int_{\rho\left(m_{1}, m_{2}, z\right)}^{\rho\left(m_{1}, m_{2}, z\right)+\mathrm{d}\rho} \frac{\mathrm{d} R}{\mathrm{d} m_{1} \mathrm{d} m_{2} \mathrm{d} z} \mathrm{d} m_{1} \mathrm{d} m_{2}  \ .
\end{align}

\subsection{Centroiding distance estimator}
\label{Est}

The conventional flux averaging estimator for GW luminosity distance is the same as the one for SNIa~\citealt{Holz:2004xx}
\begin{align}
\label{est_sim}
\left\langle\hat{d}^{2}\right\rangle=\langle\mu\rangle(d_{\mathrm{obs}})\left[\frac{1}{N} \sum_{i}^{N} d_{\mathrm{obs}, i}^{2}\right] \;,
\end{align}
where $\langle\mu\rangle(d_{\mathrm{obs}})$ is the averaged magnification given by Eq.~(\ref{ave_obs}) and $N$ stands for the number of observed GWs.
In theory, this formula is valid for a large number of events. And those redshifts have to be nearly the same and spatial distributions have to be nearly homogeneous in the observed patch of the sky. 
Thanks to the flux conservation law~\citealt{Weinberg:1976jq}, for large enough sample volume, the averaged magnification in Eq. (\ref{est_sim}) shall approach to unity. 

The above flux averaging estimator is not valid for the case of small sample volume, which is the current GW status. For single event, the lensing distribution is highly non-gaussian. This will make the flux averaging estimator sub-optimal.
For this case, we shall use the centroiding distance estimator proposed by \citealt{Hirata:2010ba}
\begin{align}
\label{eq_est}
\sum_{i=1}^{N} w\left(\ln d_{\mathrm{obs}, i}^{2}-\ln \hat{d}^{2}\right)=0 \;,
\end{align}
where $d_{\mathrm{obs},i}$ are assumed to be independent and identically distributed.
Practically, one can bin the GW data in the luminosity distance space and weight them with the lensing probability. 
This is an implicit distance estimator. 
Further, the weight function $w(x)$ is
\begin{align}
w(x)=-\frac{\mathrm{d}}{\mathrm{d} x} \ln \mathcal{P}_{d_{\mathrm{obs}}}(x, d_{\mathrm{obs}}) \;,
\end{align}
with $x=\ln\mu$, and $\mathcal{P}_{d_{\mathrm{obs}}}(x, d_{\mathrm{obs}})$ is the PDF with selection effect defined in Eq. (\ref{cal_p}).
As has been proved in \citealt{Hirata:2010ba}, the errors of this implicit estimator are always smaller than the flux averaging estimator~\citealt{Weinberg:1976jq,Holz:2004xx}. 
This is because we do input the magnification PDF information in the centroiding estimator, while we do nothing in the flux averaging method. 
This extra information help reduce the errors. 

In the limit of large $N$, which stands for the number of GWs from the identically distributed luminosity distance $d_{\mathrm{obs}}$, the maximum likelihood estimator can achieve the Fisher information errors (Cramer-Rao uncertainty bound~\citealt{Hirata:2010ba,Cusin:2020ezb}). As shown in \citealt{Hirata:2010ba}, this number is $N\simeq4$.
For this case, the uncertainty in $\ln \hat{d}^2$ would then be
\begin{align}
\sigma\left(\ln \hat{d}^{2}\right) \simeq \frac{1}{\sqrt{N I_{\ln d^{2}}^{(1)}}} \;.
\label{eq:error}
\end{align}
with
\begin{align}
I_{\ln d^{2}}^{(1)} \equiv \int \mathcal{P}_{d_{\mathrm{obs}}}(x, d_{\mathrm{obs}})\left[\frac{\mathrm{d}}{\mathrm{d} x} \ln \mathcal{P}_{d_{\mathrm{obs}}}(x, d_{\mathrm{obs}})\right]^{2} \mathrm{d} x \;,
\end{align}
and the superscript in $I_{\ln d^{2}}^{(1)}$ represents the number of GW is one.
In the rest of the paper, we will estimate the errors according to Eq.~(\ref{eq:error}) with $N=1$. 

\section{Numerical results for aLIGO/Virgo}
\label{num_results}
For simplicity, we calculate the average magnification and the luminosity distance errors by assuming a single GW detector. 
We choose the noise power spectrum of the LIGO-Livingstone detector\footnote{The noise level in LIGO-Livingstone detector is optimal among the current aLIGO/Virgo network.} 
\begin{align}
\label{noise_power}
	S_{\mathrm{fit}}(f)=\mathcal{A}_{s}\left[\left(\frac{f}{30 \mathrm{Hz}}\right)^{\alpha_{s}}+1\right] \exp \left[\beta_{s}\left(\log \left(\frac{f}{150 \mathrm{Hz}}\right)\right)^{2}\right] \ ,
\end{align}
with $\log _{10} \mathcal{A}_{s}=-46.3$, $\alpha_{s}=-7.8$ and $\beta_{s}=0.4$.

\begin{figure}
	\centering 
	\includegraphics[width=\columnwidth]{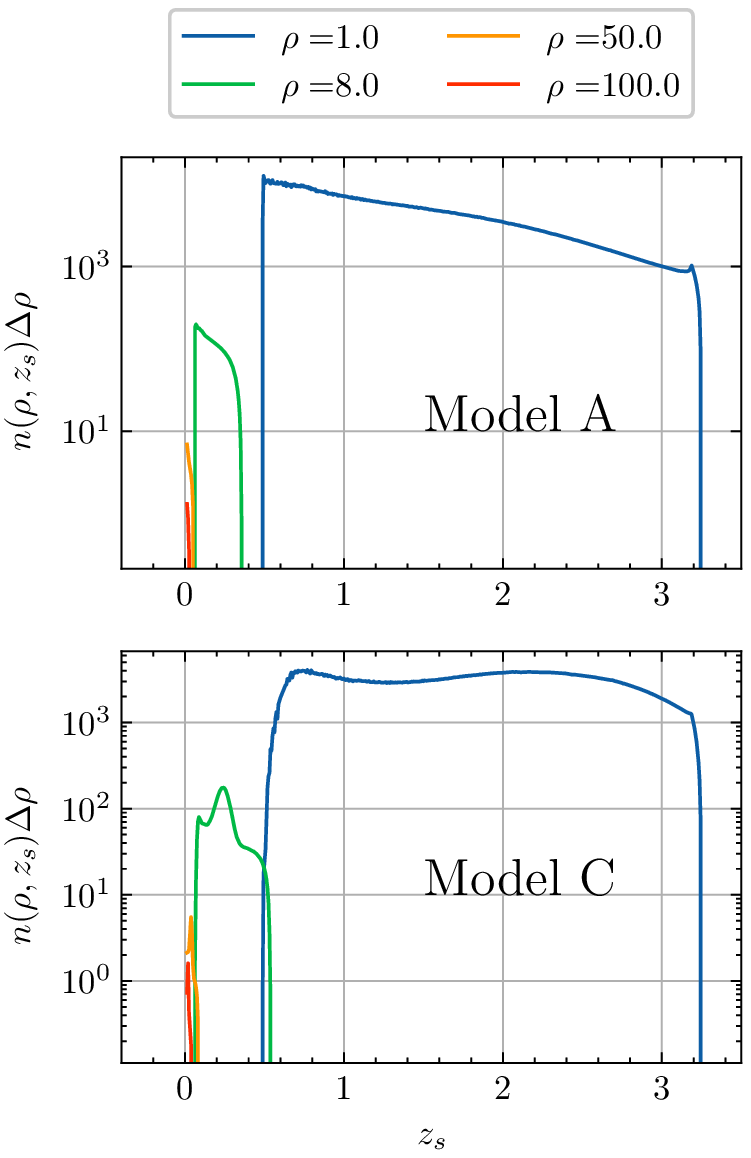}
	\caption{Y-axes are the number density of GW events per unit source redshift observed by aLIGO/Virgo in a SNR bin with the width $\mathrm{\mathrm{d}\log_{10}}\rho = 0.01$ for $\rho = 1, 8 , 50 \text{ and } 100$, respectively. 
	X-axis is the source redshift $z_s$. Upper panel shows the result for ``Model A, $\Theta = 8/5$'' and the lower panel refers to ``Model C, $\Theta = 8/5$''. The curves corresponding to different $\rho$ are shown in the legend.}
	\label{fig:N_z(rho)}
\end{figure}

In Figure~\ref{fig:N_z(rho)}, we show the number density of GW events per unit source redshift. 
The upper panel is for ``Model A, $\Theta = 8/5$'' BH mass distribution model, and the lower panel is for ``Model C, $\Theta = 8/5$''.
The vertical axes are the GW numbers density accumulated in the SNR bins with the width of $\mathrm{\mathrm{d}\log_{10}}\rho = 0.01$. 
One can see that the GWs distribution has a visible difference when applying different BH mass distribution models.
All the events whose SNR is higher than $8$ are distributed below redshift $z_s\sim 0.4$ for ``Model A'' and $z_s\sim 0.6$ for ``Model C'', because the latter has higher mass upper limit.

\begin{figure}
	\centering 
	\includegraphics[width=\columnwidth]{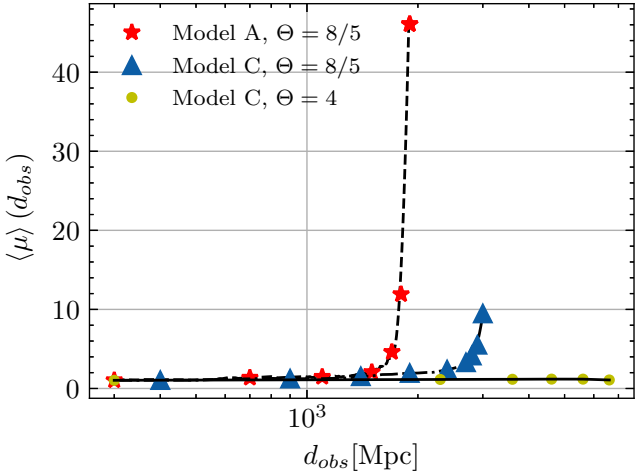}
	\caption{The average magnification of events observed by aLIGO/Virgo as a function of observed luminosity distance. The red star, blue triangle and yellow circle represent for the numerical results by using ``Model A, $\Theta = 8/5$'', ``Model C, $\Theta = 8/5$'' and ``Model C, $\Theta = 4$'', respectively. The dashed, dash-dotted and solid curves are the fitting curves for the upper three situations.}
	\label{fig:num_ave_mag_obs}
\end{figure}

We show the average magnification of events observed by aLIGO/Virgo as a function of observed luminosity distance in Figure~\ref{fig:num_ave_mag_obs}.  
One can see that the average magnification is almost unity for the observed luminosity distance $\lesssim 1600~\mathrm{Mpc}$ when using ``Model A, $\Theta = 8/5$''.
This means that, in this range, the luminosity distance estimation is unbiased for aLIGO/Virgo observation.
Moreover, for the luminosity distance $\gtrsim1600~\mathrm{Mpc}$, the average magnification suddenly increase due to the selection effect.
At $d_{\rm obs}=2000~\mathrm{Mpc}$, it will exceed $40$.
However, for ``Model C, $\Theta = 8/5$'', the higher mass upper limit makes the critical point appears at $3000~\mathrm{Mpc}$.
For $d_{\rm obs}\gtrsim3000~\mathrm{Mpc}$, the average magnification gradually increases.
Under ``Model C, $\Theta = 4$'' configuration, which is optimal for the interferometer, one can see that the average magnification keeps as unity up to $d_{\rm obs}\simeq7000~\mathrm{Mpc}$. 
Therefore, the selection effect is not important under this optimal circumstance.
We conclude that the BH mass distribution model has a sizable influence on the average magnification under the aLIGO/Virgo configuration.
Besides, the average magnification also strongly depends on the antenna pattern and BBH inclination.
~\citealt{finn1993observing} calculated the cumulative probability distribution $P\left(\Theta^2 > x^2 \right)$ of $\Theta^2$, they found that more than half (approximately $65\%$) of BBHs will have $\Theta^2 < 64/25$. For $\Theta^2 > 14.284$ the cumulative probability distribution is $0.1\%$.
Therefore, it is difficult to satisfy the ``optimal'' configuration. 
We argue that ``Model C, $\Theta = 8/5$'' is a reasonable choice.
Hence, we will focus on ``Model C, $\Theta = 8/5$'' in the following error estimation.

When $d_{\mathrm{obs}}\gtrsim 3000~\mathrm{Mpc}$, the average magnification $\langle \mu \rangle > 1$ means that we have reason to believe that these events come from a larger distance, $d_{\mathrm{true}} = \sqrt{\mu}\times d_{\mathrm{obs}}$.

\begin{figure}
	\centering
	\includegraphics[width=\columnwidth]{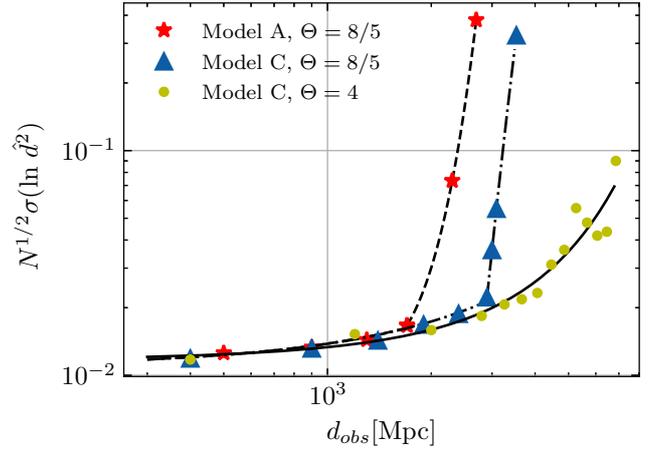}
	\caption{The estimation uncertainty on GWs observed by aLIGO/Virgo as a function of observed luminosity distance for $N=1$. The red star, blue triangle and yellow circle represent for the numerical results by using ``Model A, $\Theta = 8/5$'', ``Model C, $\Theta = 8/5$'' and ``Model C, $\Theta = 4$'', respectively. The dashed, dash-dotted and solid curves are the fitting curves for the upper three situations.}
	\label{fig:sigma_D_ligo}
\end{figure}

In Figure~\ref{fig:sigma_D_ligo}, we show the logarithmic luminosity distance estimation uncertainty $\sigma\left[{\ln (\hat{d}}/{\rm Mpc})^2\right]$ for one GW siren ($N=1$).
Hereafter, we denote as $\sigma_{\ln \hat{d}^2}$ for shorthand.  
One can see that under the aLIGO/Virgo configuration, the logarithmic luminosity distance uncertainty due to lensing magnification is below $0.02$ for the events within the observational horizon $3~\mathrm{Gpc}$ and reaches maximally at $0.3$ at $\hat{d}_{{\rm obs}}=3.5~\mathrm{Gpc}$ (for ``Model C,  $\Theta = 8/5$''). One can convert these numbers into the more intuitive relative distance errors by assuming $\sigma_{\hat{d}}/\hat{d}=\sigma_{\ln {\hat{d}}^2}/2$, 
which is $1\%$ and $15\%$ for $\hat d_{\rm obs}<3$ Gpc and $\hat d_{\rm obs}=3.5$ Gpc, respectively. 
Since the current distance errors of the aLIGO/Virgo GW events within the observational horizon are about $30\%\sim50\%$, the lensing-induced $1\%$ magnification noise is sub-dominant. 
However, for the events outside the observational horizon, namely $d_\mathrm{obs}\gtrsim 3~\mathrm{Gpc}$\footnote {Here, the observational horizon is calculated by assuming $\Theta = 8/5$ and applying $86.22-86.22~M_{\odot}$ BBHs, which is the BH mass upper-limit constrained by using O1 + O2 + O3a data. For different BH mass upper-limit and $\Theta$, the observational horizon shall be different.}, the magnification noise will dominate the error budget. 
We give a fitting formula for the uncertainty in $\ln \hat{d}^2$ for ``Model C, $\Theta = 8/5$''.
\begin{align}
\label{fit_ligo_fisher}
\begin{split}
	&\sigma_{\ln \hat{d}^2}
	=\\ &\left\{\begin{array}{ll}e^{2.28\times10^{-4}d_{\mathrm{obs}}-4.51} &  d_{\mathrm{obs}}[\mathrm{Mpc}] \leq 3000
	\\ e^{-1.24\times10^{-6}d_{\mathrm{obs}}^2 + 1.23 \times 10^{-2}d_{\mathrm{obs}} - 28.92} & 3000<d_{\mathrm{obs}}[\mathrm{Mpc}]<3500
	\end{array}\right. \;.
\end{split}
\end{align}

This result does not mean that the distance cut can be used as an absolute criteria for finding a lensed event.  
We are calculating the global statistical averaged distance cut rather than the one for a specific event. 
Especially, the values of orientation angle are adopted as the ensemble average one ($\Theta=8/5$), which can be very different from the one of a specific event.   
Hence, our result does not conflict with the lensing search result of aLIGO/Virgo \citealt{LIGOScientific:2021izm}, where the latter does not find any significant evidence for lensing.

Here, we calculate the uncertainty of a single GW siren via the Fisher information method. 
This will underestimate the errors due to the non-gaussian property of the single PDF.  
For aLIGO/Virgo, there is a high probability that we can detect a number of GWs, with $N>4$ at similar observed luminosity distance. 
In this case, the results obtained by Fisher information and Monte-Carlo Markov Chain methods will converge \citealt{Hirata:2010ba}, and the error bars will be reduced by $1/\sqrt{N}$. 

\section{Forecast for Einstein Telescope}
\label{for_et}
In this section, we forecast the distance bias and uncertainties for ET. 
We adopt a simplified sensitivity curve, which can be found on the website of ET\footnote{\url{http://www.et-gw.eu}}, and we use the ET-D version \citealt{2011CQGra..28i4013H}. 
Compared with the second generation GW detectors, such as aLIGO/Virgo, the third generation GW detectors gain at least one order of magnitude improvement in the sensitivity. 
We assume that ET can observe about $\mathcal{O}\left(10^{5}\right)$ events per year~\citealt{Vitale_2019, Chen_2020}.
We use this value to calibrate $\epsilon$ in Eq.(\ref{sfr_bbh}). 

\begin{figure}
	\centering 
	\includegraphics[width=\columnwidth]{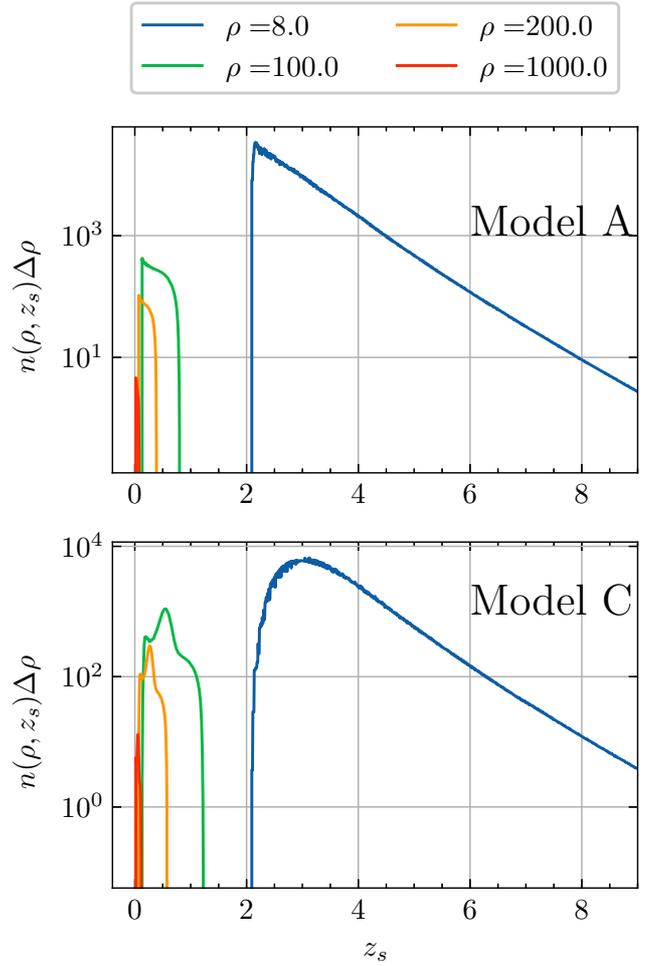}
	\caption{Y-axes are the number density of GW events per unit source redshift observed by ET in a SNR bin with width $\mathrm{dlog_{10}}\rho = 0.01$ for $\rho = 8, 100 , 200 \text{ and } 1000$, respectively. Upper panel shows the result for ``Model A, $\Theta = 8/5$'' and the lower panel refers to ``Model C, $\Theta = 8/5$''.
	X-axis is the source redshift $z_s$. The curves corresponding to different $\rho$ are shown in the legend.}
	\label{fig:ET_N_z(rho)}
\end{figure}

Figure~\ref{fig:ET_N_z(rho)} shows the distribution of GW events with fixed SNR versus the source redshifts.  
The upper panel is for ``Model A, $\Theta = 8/5$'' BBHs mass distribution model, and the lower panel is for ``Model C, $\Theta = 8/5$''.
One can identify the GWs with SNR $>8$ up to redshift $z\gtrsim 9$ or $d_\mathrm{true}\gtrsim 95~\mathrm{Gpc}$ according to the Planck cosmology.

\begin{figure}
	\centering 
	\includegraphics[width=\columnwidth]{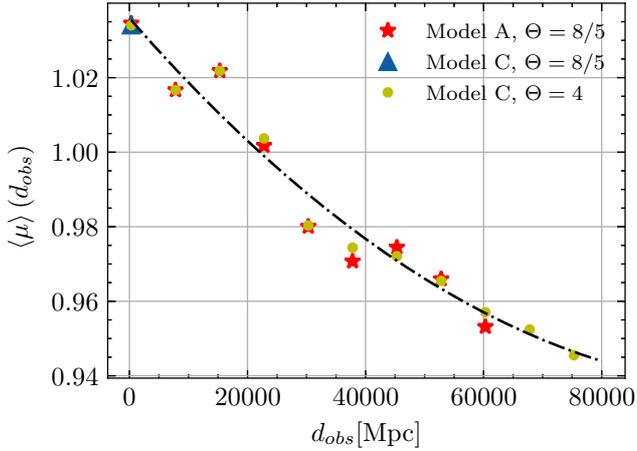}
	\caption{The average magnification of events observed by ET as a function of observed luminosity distance. The red star, blue triangle and yellow circle represent for the numerical results by using ``Model A, $\Theta = 8/5$'', ``Model C, $\Theta = 8/5$'' and ``Model C, $\Theta = 4$'', respectively. The dash-dotted curve is the fitting curve for ``Model C, $\Theta = 8/5$''.}
	\label{fig:num_ave_mag_obs_et}
\end{figure}

In Figure~\ref{fig:num_ave_mag_obs_et}, we show the average magnification of events observed by ET as a function of observed luminosity distance.  
The results of ``Model A, $\Theta = 8/5$'', ``Model C, $\Theta = 8/5$'' and ``Model C, $\Theta = 4$'' are entirely identical, implying that the selection effect is sub-dominate under the ET configuration.
Unlike aLIGO/Virgo, the average magnification monotonically decreases within $d_{\mathrm{obs}}\simeq 80~\mathrm{Gpc}$.
This is because the under-dense regime dominates the lensing effect for the GWs emitted from the deep universe.   
In detail, when $d_{\mathrm{obs}}\gtrsim 20~\mathrm{Gpc}$, the average magnification will be less than unity.
Although the average magnification crosses unity at some point, the overall deviation is not significant. 
Hence, we can conclude that the luminosity distances of GWs detected by ET are approximately unbiased, $d_{\mathrm{true}} \simeq d_{\mathrm{obs}}$.

\begin{figure}
	\centering 
	\includegraphics[width=\columnwidth]{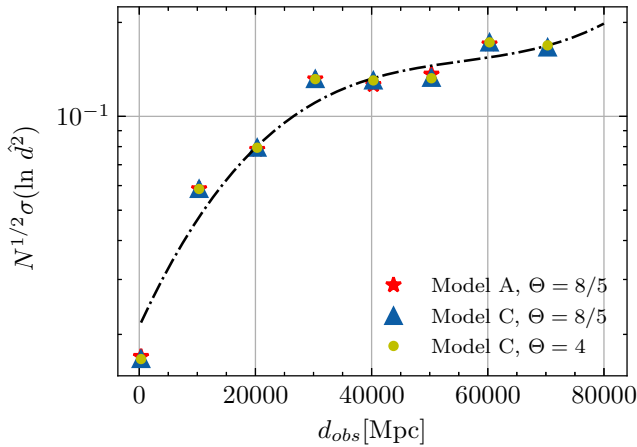}
	\caption{The estimation uncertainty on the GWs observed by ET as a function of observed luminosity distance for $N=1$. The red star, blue triangle and yellow circle represent for the numerical results by using ``Model A, $\Theta = 8/5$'', ``Model C, $\Theta = 8/5$'' and ``Model C, $\Theta = 4$'', respectively. The dash-dotted curve is the fitting curve for ``Model C, $\Theta = 8/5$''.}
	\label{fig:sigma_D_et}
\end{figure}

In Figure~\ref{fig:sigma_D_et}, we show the luminosity distance estimation uncertainty, $\sigma_{\ln \hat{d}^2}$, for a single GW siren detected by ET. 
For $d_{\mathrm{obs}}\lesssim25~\mathrm{Gpc}$, $\sigma_{\ln \hat{d}^2}$ is below $0.1$.
Above this scale, the logarithmic uncertainty will gradually reach $0.2$. 
For small errors, we can approximate $\sigma_{\hat{d}}/\hat{d}=\sigma_{\ln {\hat{d}}^2}/2$. 
We argue that a $10\%$ uncertainty in luminosity distance for $d_{\mathrm{obs}}\gtrsim25~\mathrm{Gpc}$ is not a negligible effect for ET. 
Finally, we give a fitting formula for the uncertainty in $\ln \hat{d}^2$ 
\begin{align}
\label{fit_et_fisher}
	\sigma_{\ln \hat{d}^2}= e^{9.39\times10^{-15}d_{\mathrm{obs}}^3-1.57\times{10^{-9}}d_{\mathrm{obs}}^2+9.35\times{10^{-5}}d_{\mathrm{obs}}-3.85}\;.
\end{align}

\section{Summary and discussion}
\label{sec:summary}
Due to the lensing effect, the luminosity distance of GWs obtained by the matched filtering method may not be the ``true'' distance. 
In this paper, we investigate the lensing magnification and selection effect of GWs from stellar mass BBH coalescences detected by aLIGO/Virgo O1 + O2 + O3a run. 
In detail, by using the multi-sphere ray-tracing method, we calculate the average magnification and the distance estimation uncertainties. 
Moreover, we investigate the effects of different BH mass distribution models and different orientation functions on the results. 
Finally, we forecast the same quantities for Einstein Telescope. 

In Figure~\ref{fig:num_ave_mag_obs}, we plot the average magnification as a function of observed luminosity distance which is more directly linked to observations.
Our results show that, for aLIGO/Virgo, when the luminosity distance $d_\mathrm{obs}\lesssim3~\mathrm{Gpc}$, the estimation is almost unbiased, namely $\langle\mu\rangle=1$, using the ``Model C, $\Theta = 8/5$'' configuration. 
Once $d_\mathrm{obs}\gtrsim3~\mathrm{Gpc}$, the average magnification start to deviate from unity.

The reason for this result is that the magnification probability distribution function has a sharp peak at $\mu=1$. 
For aLIGO/Virgo O1 + O2 + O3a run, on average, the observational horizon radius is about $d_\mathrm{obs}\simeq3~\mathrm{Gpc}$ when $\Theta = 8/5$. 
These imply that the GWs within this radius are probably those without intervening any foreground galaxies. 
Hence, the unlensed events dominate the expectation value of magnification. 
However, when GW events are located beyond this horizon, namely $d_\mathrm{obs}>3~\mathrm{Gpc}$, a magnification is essential to reach the qualification threshold ($\rho_\text{lim} = 8$) of aLIGO/Virgo experiments. 

The magnification estimation results depend on the BH mass distribution model for the aLIGO/Virgo case. 
For ``Model C, $\Theta = 8/5$'', the mass upper limit $m_\text{max}\sim 86~M_\odot$ is larger than the case of ``Model A, $\Theta = 8/5$'', which is $m_\text{max}=50~M_\odot$.
For the latter, the detection horizon radius is about $1.6~\mathrm{Gpc}$.
We do not know the true BH mass model, the one we adopted here is the phenomenological one obtained from the current aLIGO/Virgo data. 
Hence, the BH mass distribution model may be further modified when new observation data is available. 
However, we believe that the results will not deviate significantly from ``Model C'', because the GW frequency from BBHs with higher masses will gradually move out of the optimal frequency range of aLIGO/Virgo configuration.  

Moreover, one can see that the results also strongly depend on the orientation function $\Theta$.
Here, we consider a general configuration ($\Theta = 8/5$) and an optimal configuration ($\Theta = 4$).
We find that for ``Model C, $\Theta = 4$'' the average magnification keeps as unity up to $d_{\rm obs}\simeq7000~\mathrm{Mpc}$.
As~\citealt{finn1993observing} demonstrated that more than half (approximately $65\%$) of BBHs will have $\Theta < 8/5$ and for $\Theta > \sqrt{14.284}$ the cumulative probability distribution $P\left(\Theta^2 > 14.285 \right)$ is $0.1\%$.
Therefore, it is difficult to satisfy the ``Optimal'' configuration ($\Theta = 4$) and ``Model C, $\Theta = 8/5$'' is a reasonable choice.

We show the logarithmic luminosity distance estimation uncertainty in Figure~\ref{fig:sigma_D_ligo}.
The lensing-induced relative luminosity distance error $\sigma_{\hat{d}}/\hat{d}\sim \sigma_{\ln {\hat{d}}^2}/2$ is less than $1\%$ for $d_\mathrm{obs}\lesssim3~\mathrm{Gpc}$ by assuming ``Model C, $\Theta = 8/5$''.
Comparing with the current distance errors, namely $35\%\sim50\%$, the $1\%$ lensing-induced errors are sub-dominant. 
However, when $d_\mathrm{obs}\gtrsim3~\mathrm{Gpc}$, it can reach $\sigma_{\hat{d}}/\hat{d}=1\%\sim15\%$.
One can also see that for ``Model C, $\Theta = 4$'' case where the selection effect can be ignored, the lensing-induced relative luminosity distance error $\sigma_{\ln {\hat{d}}^2}/2$ can reach $\sim 4\%$ at $d_{\mathrm{obs}}\simeq 7~\mathrm{Gpc}$.

For Einstein Telescope, we find that when the luminosity distance $d_{\mathrm{obs}}\gtrsim 25~\mathrm{Gpc}$, the average magnification will be less than $1$.
The under-dense contributions domain the expectation. 
Furthermore, due to the robustness of ET, we are able to identify almost all the GWs within its observational horizon $d_\mathrm{true}\simeq95~\mathrm{Gpc}$ or $z_s\sim 9$.
Hence, the average magnification is only slightly different from $1$.
We also estimate the luminosity distance uncertainty caused by lensing.
For $d_{\mathrm{obs}}\gtrsim25~\mathrm{Gpc}$, the relative luminosity distance estimation error, $\sigma_{\hat{d}}/\hat{d}$, will gradually reach $10\%$.
Unlike the aLIGO/Virgo case, this scatter is not due to the selection effect. It purely comes from the unavoidably accumulated lensing magnification. 
Finally, we need to emphasize that the different BH mass distribution models and common orientation functions shall not bring any sizable differences in the magnification estimation under the ET's configuration.

\section*{Acknowledgements}
We thank Ruth Durrer, Giulia Cusin and Xiaoyue Cao for useful discussions. 
We also thank an anonymous referee for the discussion on the aLIGO/Virgo lensing search result.
This work is supported by the National Natural Science Foundation of China Grants No. 11973016, No. 11690023, No. 11653003 and No.11903082. 


\section*{Data availability}
The data underlying this article will be shared on reasonable request to the corresponding author.

\bibliographystyle{mnras}
\bibliography{GW_Lens} 



\appendix

\section{Selection effect}
\label{GW_w_redshift}

In the appendix, we demonstrate the selection effect for aLIGO/Virgo and Einstein Telescope.
\subsection{aLIGO/Virgo}

\begin{figure}
	\centering
	\includegraphics[width=\columnwidth]{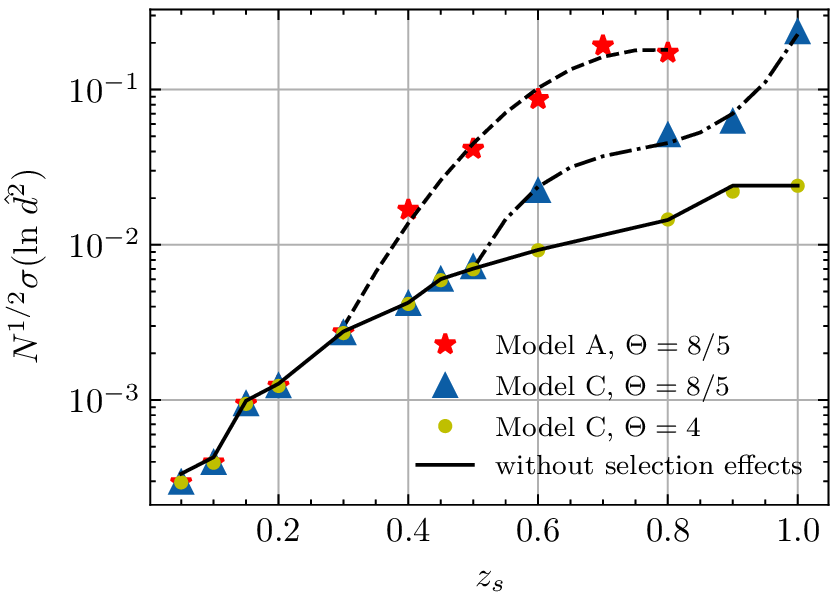}
	\caption{The estimation uncertainty on the GWs observed by aLIGO/Virgo as a function of $z_s$ for $N=1$. The red star, blue triangle and yellow point represent the numerical results by using ``Model A, $\Theta = 8/5$'', ``Model C, $\Theta = 8/5$'' and ``Model C, $\Theta = 4$'', respectively. The dashed, dash-dotted and solid curves are the fitting curves for the upper three situations.}
	\label{fig:sigma_D_ligo_z}
\end{figure}

Figure~\ref{fig:sigma_D_ligo_z} shows the estimation uncertainty $\sigma_{\ln \hat{d}^2}$ for a single GW siren as a function of redshift. 
One can see that the selection effect is not visible when $z\lesssim1$ ($z\lesssim0.5$, $z\lesssim0.3$) for ``Model C, $\Theta = 4$'' (``Model C, $\Theta = 8/5$'', ``Model A, $\Theta = 8/5$'').
We give a fitting formula for the uncertainty in $\ln \hat{d}^2$ with the selection effect for ``Model C, $\Theta = 8/5$''
\begin{align}
\label{fit_ligo_fisher_z}
	\sigma_{\ln \hat{d}^2}
	=\left\{\begin{array}{ll} e^{83.23z^3-187.73z^2 + 142.89z - 39.87} & \text { for }  0.5 \le z \le 1 
	\\ 0.019\left[\frac{1-(1+z)^{1.36}}{-1.36}\right]^{1.73} & \text { for } z<0.5 \end{array}\right. \;.
\end{align}
In addition, we give the fitting formula for the case without the selection effect
\begin{align}
\label{fit_ligo_fisher_z_wo}
	\sigma_{\ln \hat{d}^2} = 0.019\left[\frac{1-(1+z)^{1.36}}{-1.36}\right]^{1.73} ,\text { for } z<1 \;.
\end{align}

\subsection{Einstein Telescope}

\begin{figure}
	\centering 
	\includegraphics[width=\columnwidth]{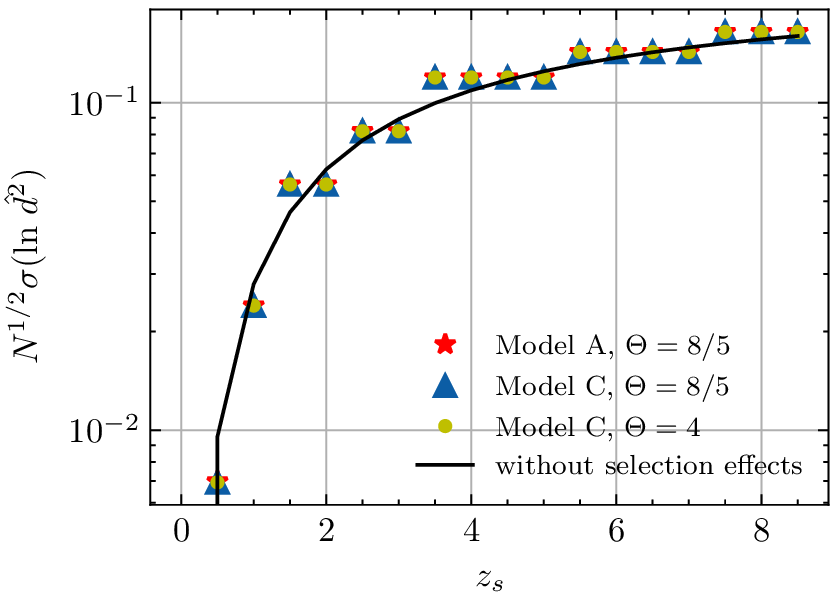}
	\caption{The estimation uncertainty on the GWs observed by ET as a function of $z_s$ for $N=1$. The solid curve is the fitting formula in Eq.~(\ref{fit_et_fisher_z_wo}). The red star, blue triangle and yellow point represent the numerical results by using ``Model A, $\Theta = 8/5$'', ``Model C, $\Theta = 8/5$'' and ``Model C, $\Theta = 4$'', respectively.}
	\label{fig:sigma_D_et_z}
\end{figure}

In Figure~\ref{fig:sigma_D_et_z}, we show the estimation uncertainty $\sigma_{\ln \hat{d}^2}$ for a single siren as a function of $z_s$ in ET.
One can see that the selection effect is not visible for these three configurations.
We give the fitting formula for the case without the selection effect
\begin{align}
\label{fit_et_fisher_z_wo}
	\sigma_{\ln \hat{d}^2} = 0.12\left[\frac{1-(1+z)^{-0.72}}{0.72}\right]^{2.44} \;.
\end{align}


\bsp	
\label{lastpage}
\end{document}